\documentclass[12pt]{article}

\usepackage{graphicx}

\usepackage[english]{babel}
   \usepackage[cp1251]{inputenc}

\usepackage{colortbl}
\usepackage{float}

\def\ppluss{\subset\!\!\!\!\!\!~+}

\begin{document}

\begin{center}
{\large \bf
The contraction hypothesis of the gauge group of the Standard Model
and LHC experimental data
 }
\end{center}

\begin{center}
Nikolai~A.~Gromov  \\
Institute of Physics and Mathematics, Komi Science Centre UrB RAS, \\
Kommunisticheskaya st. 24, Syktyvkar 167 982, Russia \\
E-mail: gromov@ipm.komisc.ru
\end{center}

\begin{center}
     {\bf  Abstract }
\end{center}

Within the framework of the contraction hypothesis of the gauge group of the Standard Model
the behavior of the amplitude of the dominant Higgs boson production process in the four-lepton decay with increasing temperature $T$ is analyzed.
It is shown that the modified process breaks down into a number of channels depending on the contribution of the color components in the loop of virtual quarks, leading to the creation 
of the Higgs boson.
The dependence on $T$ of the cross section of each channel is found.
Comparison with LHC data on Higgs boson creation  cross sections at energies (temperatures) of 7, 8, 13, and 14 TeV
showed that
the hypothesis about the contraction of the gauge group of the Standard Model does not contradict these data.

\vspace{3mm}





\section{Introduction}


The theory of elementary particles and their interactions - the Standard Model - has been confirmed by experiments at the Large Hadron Collider (LHC) carried out in the last decade.
The theoretically predicted scalar Higgs boson was discovered, data were obtained on the cross section for the production of the Higgs boson at different energies of 7, 8, 13, and 14 TeV.
The latter value represents the practical limit of the energies available to modern accelerators.
The emergence of new accelerating machines with higher energies is unlikely due to the colossal costs of their construction.
All this stimulates interest in the theoretical study of the properties of particles and the processes of their interaction at high energies.
The basis for such a study, if we want to stay on solid scientific ground, is the Standard Model and its modifications, in particular, the behavior
 at high energies.

The standard model includes
the electroweak model \cite{R-99}, describing electro\-magnetic and weak interactions of particles, and quantum chromodynamics (QCD) \cite{Em-2007}, which describes the strong interactions of quarks.
It is a gauge theory based on the gauge group $SU(3)\times SU(2)\times U(1)$, which is a direct product of simple groups.
The strong interactions of quarks are described by quantum chromodynamics with a gauge group of $SU(3)$ and a characteristic temperature of $0.2$ GeV.
In the electroweak model with the gauge  group $SU(2)\times U(1)$, the group $SU(2)$ is responsible for weak interactions with a characteristic temperature of $100$ GeV, whereas the group $U(1)$ is associated with long-range electromagnetic interactions. 
Due to the zero mass of the photon -- the carrier of this interaction -- its characteristic temperature extends to the "infinite" $\,$ Planck energy $10^{19}$ GeV.

Starting from the observation of characteristic energies,
we propose \cite{Gr-2016,Grom-2020,Gr-2020} a new hypothesis in particle physics: {\bf the gauge group of the Standard Model becomes simpler with increasing energy (temperature) of the universe.}
We assume that with an increase in energy (temperature), a simpler gauge  group of the Standard Model is obtained using
{\bf contractions of the group $SU(3)\times SU(2)\times U(1)$,
whose contraction parameter decreases as the temperature of the Universe increases.}
Since the average energy  $T$ of a hot Universe is related to its age \cite{GoR-11,L-1990},
the contraction parameter   $\varepsilon  \sim  T^{-q},\; q>0$ 
tends to zero at $T\rightarrow\infty$.

The operation of contraction (or limiting transition) groups is well known in physics \cite{IW-53}.
In particular, it transforms a simple group into a non-simple one.
The concept of contraction has been extended \cite{Gr-12} to algebraic structures such as quantum groups, supergroups, as well as to fundamental representa\-tions of unitary groups that are directly related to the Standard Model.
For a symmetric physical system, the contraction of the symmetry group means a transition to one or another limiting state of the system.
In the case of a complex physical system, which is the Standard Model, the study of limit states at certain limit values of physical parameters opens up the possibility to better understand the behavior of the system as a whole.

Since the change of the gauge  group during the contraction occurs conti\-nuously, including at the very beginning of the limit transition at values of the parameter $\varepsilon$ near unity, we can try to catch the effect of the contraction effect by comparing the data obtained at the LHC on the cross section of the Higgs boson at different energies with the theoretical dependence of the cross section on the temperature of the Universe.
In this paper, we analyze the dominant mechanism of the creation  and registration of Higgs bosons at the LHC in a four-lepton process,
considering the temperature dependence of the corresponding Feynman diagram.
The modification of Feynman diagrams and the dependence of the cross sections of electroweak processes on the contraction parameter
has been previously  considered in \cite{G-2020,Gr-2021}.

\section{Gauge group contraction  and field trans\-for\-mation} 

The electroweak model combining electromagnetic and weak interactions is a gauge theory with a gauge group $SU(2)\times U(1)$ acting in the space 
${\bf C}_2$ of the fundamental representation of the group $SU(2)$.
Moreover, the points of the space ${\bf C}_2$ represent four (or eight, if antiparticles are taken into account) component spinors, and
the vectors describe leptons:
$$
\left(
\begin{array}{c}
	\nu_{e} \\
	e  
\end{array} \right), \;
\left(
\begin{array}{c}
	\nu_{\mu} \\
	\mu  
\end{array} \right), \;
\left(
\begin{array}{c}
	\nu_{\tau} \\
	\tau  
\end{array} \right),\; 
$$
where $e$ is an electron, $\mu$ is a muon, and $\tau$ is a lepton,
$\nu_{e}$, $\nu_{\mu}$, $\nu_{\tau} $ -- corresponding neutrinos,
as well as three generations of quarks:
$$
 \left(
\begin{array}{c}
	u \\
	d 
\end{array} \right), \;
\left(
\begin{array}{c}
	c \\
	s 
\end{array} \right), \;
\left(
\begin{array}{c}
	t \\ 
	b 
\end{array} \right) . 
$$ 
In the future, we will consider only the first generations of leptons and quarks.
The matrix elements of the group $SU(2)$ define the gauge  bosons:
$\gamma\;$ -- photon,
$Z\;$ -- neutral weak boson,
$W^{\pm}\;$ -- charged weak bosons.
We introduce the contracted group
$SU(2;\varepsilon)$ and the corresponding fundamental representation space $\bf{C}_2(\varepsilon)$\cite{Gr-12}
by a coordinated change of the elements of the group $SU(2)$ and the component of the space ${\bf C}_2$ of the form
$$ 
\left( \begin{array}{c}
	z'_1 \\
{\varepsilon}	z'_2
\end{array} \right)
=\left(\begin{array}{cc}
	\alpha & {\varepsilon}\beta   \\
-{\varepsilon}\bar{\beta}	 & \bar{\alpha}
\end{array} \right)
\left( \begin{array}{c}
	z_1 \\
{\varepsilon}	z_2
\end{array} \right), 
$$
\begin{equation}
\det u(\varepsilon)=|\alpha|^2+{\varepsilon^2}|\beta|^2=1, \quad u(\varepsilon)u^{\dagger}(\varepsilon)=1.
\label{5}
\end{equation}
%
Our approach is based on the action of matrices with elements depending on the contraction
parameter $\varepsilon$ on vectors whose components also depend on this parameter.

Replacing $\beta\rightarrow {\varepsilon}\beta$ induces a transformation of Lie algebra generators $su(2)$:
$T_1 \rightarrow \varepsilon T_1$, $T_2 \rightarrow \varepsilon T_2$, $T_3 \rightarrow T_3$.
In the limit $\varepsilon\rightarrow 0$, the simple algebra $su(2)$ acquires the structure of a semidirect sum
$t_2 \ppluss t_1$ of a commutative ideal $t_2=\{T_1, T_2\}$ and a one-dimensional subalgebra $t_1=\{T_3\} $.
Representation space $\bf{C}_2(\varepsilon)$ is stratified in this limit into a one - dimensional base
$\{z_1\}$ and a one-dimensional layer $\{z_2\} $.

Since gauge fields take values in Lie algebra, it is possible to transform gauge  fields instead of generators.
For standard gauge fields \cite{R-99}, this transformation has the form
\begin{equation}
W_{\mu}^{\pm} \rightarrow {\varepsilon}W_{\mu}^{\pm}, \quad Z_{\mu} \rightarrow Z_{\mu},\quad A_{\mu} \rightarrow A_{\mu}. 
\label{g15-b}
\end{equation}
The left lepton 
$
L_l= \left(
\begin{array}{c}
	\nu_l\\
	e_{l}
\end{array} \right)\;
$
and quark  
$
Q_l= \left(
\begin{array}{c}
	u_l\\
	d_{l}
\end{array} \right)
$
 fields are $SU(2)$-doublets (vectors), so their components are transformed as components of the vector $z$, namely:
\begin{equation}
 	 e_{l} \rightarrow \varepsilon e_{l},  \quad  d_{l} \rightarrow \varepsilon d_{l}, \quad
 	 \nu_l \rightarrow \nu_l, \quad  	u_l \rightarrow u_l.
\label{6}
\end{equation}
The right lepton and quark fields are $SU(2)$-singlets (scalars) and therefore do not change.

The group $U(1)$ consisting of multiplications of vectors from the representation space by a unimodular complex number
\begin{equation}
U(1): \; \vec{z}\,'=e^{i\omega/2}\vec{z},\quad \omega\in \mathbf{R}.
\label{eq1qq}
\end{equation}
is not transformed during contraction.
The same is true for its gauge  boson, the photon $\gamma$.

In the mechanism of spontaneous symmetry breaking, by which the masses of vector bosons and other particles of the electroweak model are generated, one of the basic states of the bosonic Lagrangian 
is selected as the vacuum of the model and then small perturbations relative to this vacuum are considered
$  
\phi=\left(\begin{array}{c}
	0  \\
	v+H(x)
\end{array} \right). \;
$
Therefore, the Higgs boson field $H(x) $, the constant $v$ and the particle masses $m_p$ depending on it are multiplied by the contraction parameter:
\begin{equation}
 	H  \rightarrow \varepsilon H,  \quad  v \rightarrow \varepsilon v, \quad 
 m_p \rightarrow \varepsilon m_p, 
\label{7}
\end{equation}
where $p={H}, W, Z, e, u, d$.

Substitutions (\ref{g15-b})--(\ref{7}) in the Lagrangian of the electroweak model give the transformed
Lagrangian of the electroweak model with a contracted gauge group, which takes the form
\begin{equation}
{L}_{EWM}(\varepsilon)= L_{\infty} + \varepsilon L_{1} 
 + \varepsilon^2 L_{2} + \varepsilon^3 L_{3}  + \varepsilon^4 L_{4}.
\label{8}
\end{equation}
The explicit form of the terms $L_k$ can be found in the monograph \cite{Gr-2020}.
With $\varepsilon\rightarrow 0 $ terms with higher powers
$\varepsilon$ contribute less to the Lagrangian than terms with low degrees.
Thus, a modified electroweak model with an increase in temperature up to "infinite"$\,$ $10^{19} $ GeV demonstrates five stages of behavior,
which differ in the degrees of the contraction parameter,
which largely removes the problem of hierarchies \cite{Em-2007}.

Quantum chromodynamics is a gauge theory with a group $SU(3)$ acting in a three-dimensional complex space ${\bf C}_3$
color states of quarks
$ q=(q_1,q_2,q_3)^t\equiv (q_R,q_G,q_B)^t \in {\bf C}_3 $,
where $q(x)$ denotes the quark fields $q=u, d, s, c, b, t$, and the indices $R$ (red), $G$ (green), $B$(blue) denote the color degrees of freedom \cite{Em-2007}.
Contraction of the  QCD gauge  group  $SU(3;\varepsilon)$ is determined by the action of the group
$q'({\varepsilon})= U({\varepsilon} )q({\varepsilon}) $ of the form
\begin{equation}
\left(\begin{array}{c}
 q'_{1}\\
{\varepsilon} q'_{2} \\
{\varepsilon^2} q'_{3}
 \end{array}
 \right)=
\left(\begin{array}{ccc}
 u^{11}  &{\varepsilon} u^{12} &{\varepsilon^2} u^{13} \\
 {\varepsilon} u^{21} & u^{22} & {\varepsilon} u^{23} \\
 {\varepsilon^2} u^{31} & {\varepsilon} u^{32} & u^{33}
 \end{array}
 \right)
 \left(\begin{array}{c}
 q_{1}\\
{\varepsilon} q_{2} \\
{\varepsilon^2} q_{3}
 \end{array}
 \right)
 \label{10}
\end{equation}
 in the color space $\mathbf{C}_3({\varepsilon})$ of the fundamental representation at ${\varepsilon}\rightarrow 0$.
 At the same time, the degenerate Hermitian form remains invariant
\begin{equation}
 q^{\dagger}(\varepsilon )q(\varepsilon )= \left|q_1\right|^2+  \varepsilon^2\left|q_2\right|^2 +  \varepsilon^4 \left|q_3\right|^2
 \label{q8}
\end{equation}
in the double-layered space $\mathbf{C}_3({\varepsilon}=0)$ with a base $\{q_1\}$,
and a two-dimensional layer $\{q_2,q_3\}$, which in turn is layered on the base $\{q_2\}$ and layer $\{q_3\}$.
The structure of the contracted algebra $su(3;\varepsilon)$ can be represented as
    $$
   su(3;\varepsilon=0)=
   $$
   $$
   =\Big( t_2\{\lambda_3,\lambda_8\} \ppluss t_2\{\lambda_1,\lambda_2\}\Big) \ppluss
   t_4\{\lambda_4,\lambda_5,\lambda_6,\lambda_7\}=
   $$
  \begin{equation}
  =\Big(t_2\{\lambda_3,\lambda_8\} \ppluss t_2\{\lambda_6,\lambda_7\}\Big) \ppluss
   t_4\{\lambda_1,\lambda_2,\lambda_4,\lambda_5\},
  \label{s3}
 \end{equation}  
 where $ \lambda_i,\; i=1,\ldots,8\; $ are  the Gell-Mann matrix  \cite{GKK-2020}.

Transition from the classical group $SU(3)$ and the complex space ${\bf C}_3$ to the group $SU(3;\varepsilon)$ and the space
${\bf C}_3(\varepsilon)$ is carried out by substitutions in the Lagrangian of the standard QCD of gluon and quark fields of the form
$$
u^{12(21)}_{\mu} \rightarrow \varepsilon u^{12(21)}_{\mu},\quad
u^{23(32)}_{\mu}\rightarrow \varepsilon u^{23(32)}_{\mu},\quad
u^{13(31)}_{\mu}\rightarrow \varepsilon^2 u^{13(31)}_{\mu},
$$
\begin{equation}
q_1 \rightarrow q_1, \quad
q_2\rightarrow \varepsilon  q_2,\quad q_3\rightarrow \varepsilon^2  q_3.
 \label{q9}
\end{equation}
Diagonal gauge  fields (gluons) $u^{11}_{\mu}, u^{22}_{\mu}, u^{33}_{\mu}$
are not transformed  in this case.

The substitutions (\ref{q9}) lead to the quark part of the QCD Lagrangian of the form
\begin{equation}
{\cal L}_q(\varepsilon )= L_q^{0}  
+ \varepsilon^2 L_q^{(2)} + 
\varepsilon^4 L_q^{(4)}, 
 \label{q11-C}
\end{equation}
where 
\begin{equation} 
L_q^{0} =
\sum_q   \Biggl\{ i\bar{q}_1\gamma^\mu\partial_{\mu} q_1 - m_q\left|q_1\right|^2
+ 
\frac{g_s}{2} \left|q_1\right|^2 \gamma^\mu  u^{11}_{\mu} 
\Biggr\},
 \label{Q-0}
\end{equation}
$$
L_q^{(2)}= \sum_q  \Biggl\{
i\bar{q}_2\gamma^\mu \partial_{\mu} q_2 - m_q\left|q_2\right|^2
+  
$$
\begin{equation}
   +\frac{g_s}{2}\biggl(\left|q_2\right|^2 \gamma^\mu u^{22}_{\mu}
+q_1\bar{q}_2\gamma^\mu u^{21}_{\mu}  
+ \bar{q}_1q_2\gamma^\mu  u^{12}_{\mu} 
\biggr)
 \Biggr\},
 \label{Q-2}
\end{equation}
$$
L_q^{(4)}= \sum_q \Biggl\{
i\bar{q}_3\gamma^\mu \partial_{\mu} q_3 - m_q\left|q_3\right|^2 
+
 \frac{g_s}{2}\biggl(\left|q_3\right|^2 \gamma^\mu u^{33}_{\mu} +
$$
\begin{equation}
+q_1\bar{q}_3\gamma^\mu  u^{31}_{\mu} 
+ \bar{q}_1q_3\gamma^\mu u^{13}_{\mu} + 
+ q_2\bar{q}_3\gamma^\mu u^{32}_{\mu} 
+ \bar{q}_2q_3\gamma^\mu u^{23}_{\mu} 
\biggr)\Biggr\}.
 \label{Q-4}
\end{equation}
The gluon part of the Lagrangian is very cumbersome and we do not present it here.

\begin{figure}[H] 
\begin{center}
\includegraphics[width=90mm]{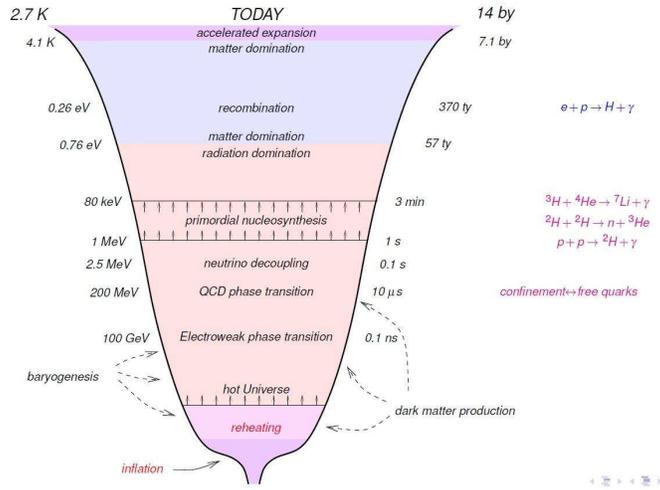}
\end{center}
      \caption{History of the Universe \cite{GoR-11,L-1990}.}
 \label{fig1}   
\end{figure}

Complete Lagrangian of modified QCD
with a contracted gauge group 
can be written as a power expansion of the contraction parameter
\begin{equation}
{ L}_{QCD}(\varepsilon)={ L}^{0} 
+ \varepsilon^2 { L}^{(2)} 
+ \varepsilon^4 { L}^{(4)} 
 + \varepsilon^6 { L}^{(6)} + \varepsilon^8 { L}^{(8)}.
 \label{q11}
\end{equation}
The explicit form of the terms $L^{(k)}$ is given in the monograph \cite{Gr-2020}.
As a result, for QCD we  have five stages of behavior with an increase in the temperature of the Universe,
i.e., when moving backwards in time to the moment of its birth as a result of the Big Bang 
(fig.\ref{fig1}).


\section{Higgs boson creation  cross section in LHC experiments}

\begin{figure}[H] 
\begin{center}
\includegraphics[width=70mm]{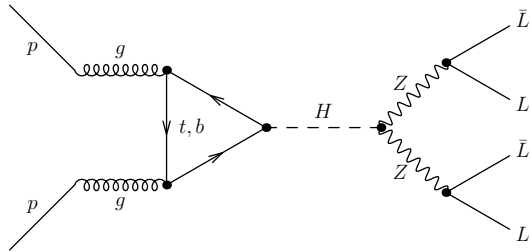}
\end{center}
   \caption{Diagram of Higgs boson production in a four-lepton process.}
 \label{fig2}   
\end{figure}

The dominant mechanism of the creation  and registration of Higgs bosons at the LHC is described by the Feynman diagram shown in Fig.\ref{fig2}.

In the colliding proton
beams, two gluons binding quarks into hadrons combine into a loop of virtual quarks ($t$ or $b$ type),
which gives rise to the Higgs boson $H$, which then decays into a pair of neutral $Z$ bosons. Subsequently, each of the $Z$ bosons decays
into a pair of charged leptons $L$ (electrons or muons).
Simultaneous registration of four leptons is an indicator of the birth of the Higgs boson.

\begin{figure}[H] 
\begin{center}
\includegraphics[height=40mm]{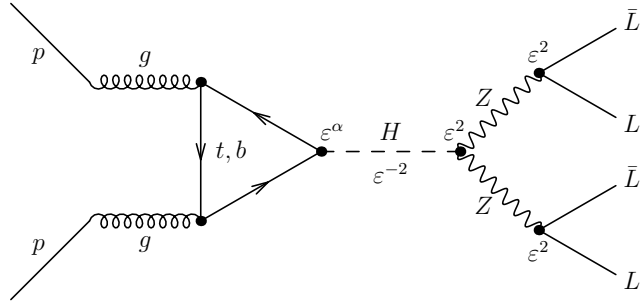}
\end{center}
      \caption{Modified diagram of Higgs boson production in a four-lepton process.}
    \label{fig3}
\end{figure}

From the Lagrangian of the modified electroweak model (\ref{8}) and field transformations
\begin{equation}
t \rightarrow t, \quad  b \rightarrow {\varepsilon} b, \quad Z \rightarrow Z, \quad H \rightarrow {\varepsilon} H
 \label{q12}
\end{equation}
taking into account that the propagator is the inverse operator to the equation of a free particle, i.e.
if the Higgs boson equation is multiplied by $\varepsilon^2$, then its propagator is by $\varepsilon^{-2}$,
we get the transformed diagram shown in Fig.\ref{fig3}.
After calculating the contract multipliers $\varepsilon$ characterizing the right electroweak part of the diagram,
it can be represented in a form that depends only on the strong interactions of quarks.
\begin{figure}[H] 
\begin{center}
  \includegraphics[height=40mm]{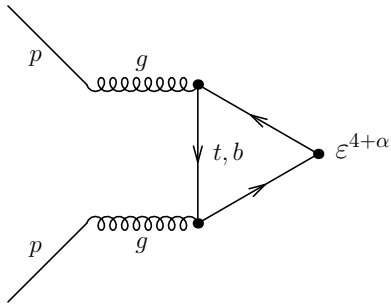}    
\end{center}
      \caption{Higgs boson production diagram dependent on strong quark interactions.
                Here $\alpha =1$ for the $t$-quark and $\alpha =2$ for the $b$-quark. }
    \label{fig4}
\end{figure} 

Diagram Fig.\ref{fig4} is modified by contracting the group $SU(3;\varepsilon) $.
If in the initial loop of quarks their components and gluons are equal, then after the contraction
there is a "splitting" of the processes of formation of Higgs bosons, associated with a different dependence
of the colors (components) of quarks
on ${\varepsilon}=(A T^{-1})^q, q>0$, where
$A=4\cdot 10^{-4}$ if $T$ is measured in GeV
and $A=4\cdot 10^{-7}$ if $T$ is measured in TeV.
Using (\ref{q11-C})--(\ref{Q-4}), we get 9 different loops of virtual quarks, multiplied by the contraction factor in different powers.

\begin{figure}[H] 
\begin{center}
\includegraphics[height=45mm]{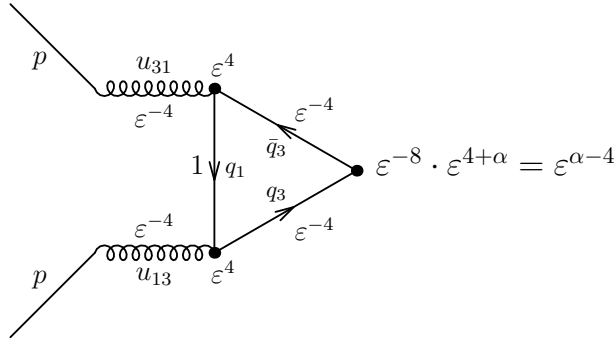}  
\end{center}
   \caption{A loop of virtual quarks with components $q_1, q_3$ and an antiquark with component $ \bar{q}_3 $. Channel amplitude $M_{31}(\varepsilon)$.}
    \label{fig5}
\end{figure}

\begin{figure}[H] 
\begin{center}
\includegraphics[height=45mm]{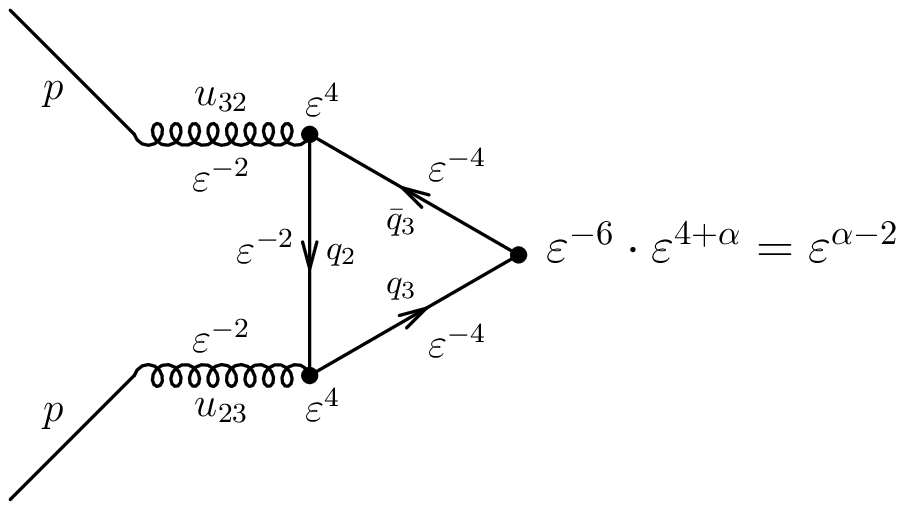}  
\end{center}
   \caption{A loop of virtual quarks with components $q_2, q_3$ and an antiquark with component $ \bar{q}_3 $. Channel amplitude $M_{32}(\varepsilon)$.}
    \label{fig5a}
\end{figure}

One loop (Fig.\ref{fig5}) gives the multiplier $ \varepsilon^{\alpha -4} $.
Another loop (Fig.\ref{fig5a}) gives the $ \varepsilon^{\alpha -2} $ multiplier.

\begin{figure}[H] 
\begin{center}
\includegraphics[height=35mm]{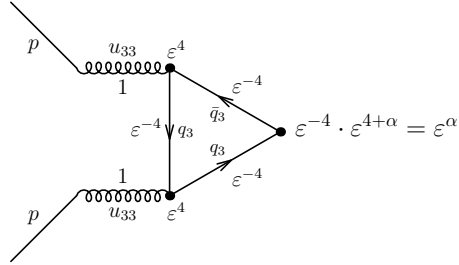}    
\end{center}
     \caption{A loop of virtual quarks and antiquarks with the third  component. Channel amplitude $M_{33}(\varepsilon)$.}
    \label{fig6-0}
\end{figure}
\begin{figure}[H] 
\begin{center}
\includegraphics[height=35mm]{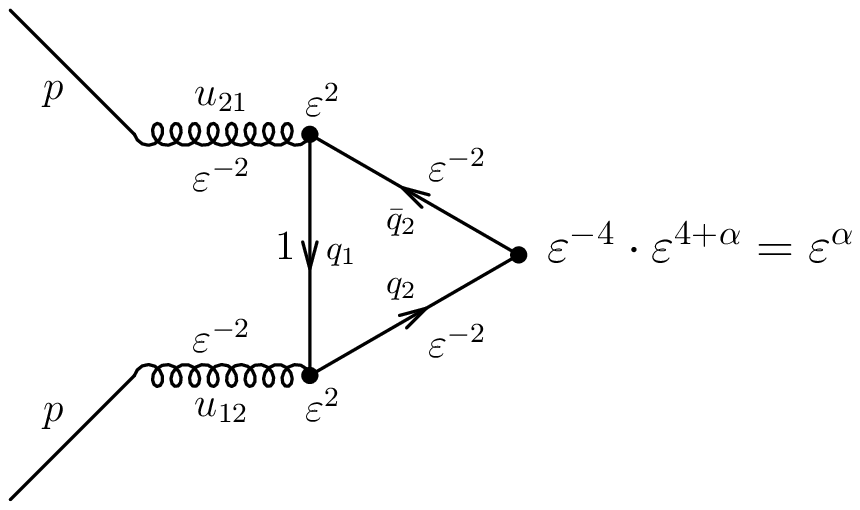}    
\end{center}
     \caption{A loop of virtual quarks with components $q_1, q_2$ and an antiquark with component $ \bar{q}_2 $. Channel amplitude $M_{21}(\varepsilon)$.}
    \label{fig6-1}
\end{figure}
\begin{figure}[H] 
\begin{center}
\includegraphics[height=35mm]{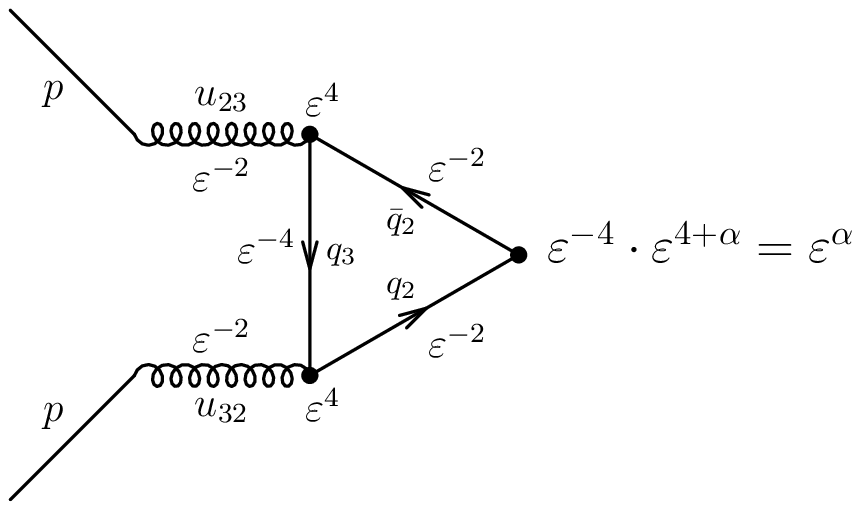}    
\end{center}
     \caption{A loop of virtual quarks with components $q_3, q_2$ and an antiquark with component $ \bar{q}_2 $. Channel amplitude $M_{23}(\varepsilon)$.}
    \label{fig6-2}
\end{figure}
\begin{figure}[H] 
\begin{center}
\includegraphics[height=37mm]{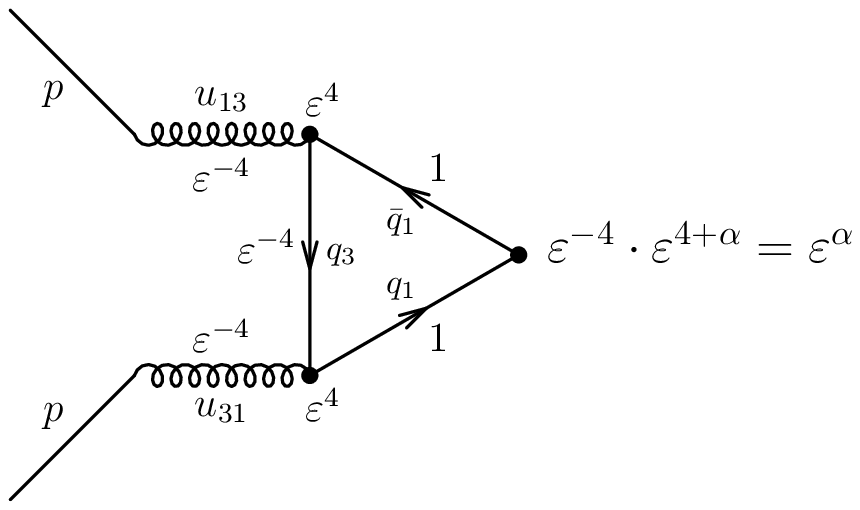}    
\end{center}
     \caption{A loop of virtual quarks with components $q_3, q_1$ and an antiquark with component $ \bar{q}_1 $. Channel amplitude $M_{13}(\varepsilon)$.}
    \label{fig6-3}
\end{figure}
\begin{figure}[H] 
\begin{center}
\includegraphics[height=37mm]{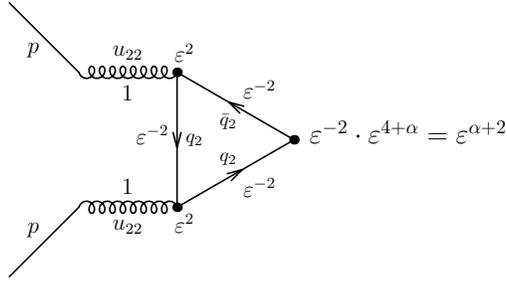}    
\end{center}
     \caption{A loop of virtual quarks and antiquarks with the second component.  Channel amplitude $M_{22}(\varepsilon)$.}
    \label{fig6-4}
\end{figure}
\begin{figure}[H] 
\begin{center}
\includegraphics[height=37mm]{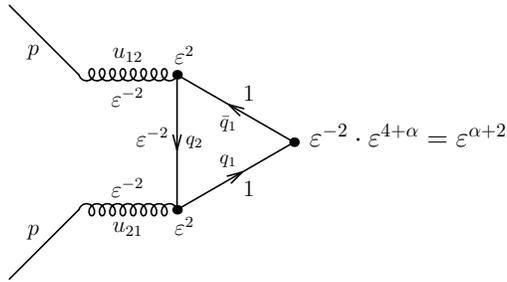}    
\end{center}
     \caption{A loop of virtual quarks with components $q_2, q_1$ and an antiquark with component $ \bar{q}_1 $. Channel amplitude $M_{12}(\varepsilon)$.}
    \label{fig6-5}
\end{figure}
\begin{figure}[H] 
\begin{center}
\includegraphics[height=37mm]{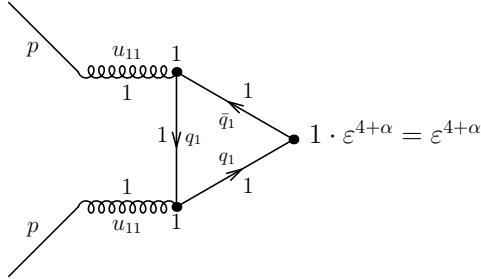}    
\end{center}
     \caption{A loop of virtual quarks and antiquarks with the first component. Channel amplitude $M_{11}(\varepsilon)$.}
    \label{fig6}
\end{figure}

Four loops  (Fig.\ref{fig6-0}--Fig.\ref{fig6-3}) result in a factor $ \varepsilon^{\alpha } $.
Two loops  (Fig.\ref{fig6-4}, Fig.\ref{fig6-5}) have a multiplier  $ \varepsilon^{\alpha +2} $.
One loop  (Fig.\ref{fig6}) of virtual quarks has a multiplier $ \varepsilon^{\alpha +4} $.

Thus, the amplitudes of the processes of the Higgs boson generation are multiplied by the contraction parameter in various degrees
depending on which color components of virtual quarks are involved in its formation.
This dependence is described by the expression
\begin{equation}
M_{ik}(\varepsilon) =  \varepsilon^r M_{ik}^0,\quad i,k=1,2,3,
 \label{q13}
\end{equation}
 where $r= -3,-1,1,3,5$ for $t$-quark and $r= -2,0,2,4,6$ for $b$-quark.
Here $M_{ik}^0=M_{ik}(\varepsilon=1)$ is the initial amplitude of an unmodified process with quark components $q_i$ and $q_k$. 
The cross section of the process is proportional to the square of the amplitude $\sigma_{ik} = |M_{ik}|^2$. 
Since the parameter $ \varepsilon = (AT^{-1})^q $
is small, then the main contribution to the total cross-section $\sigma^{tot}$ with increasing temperature $T$ comes from channels proportional to
contraction parameter with negative powers,
i.e.
\begin{equation}
\sigma_t(T) = \sigma_t^0\cdot \varepsilon^{-p} = \sigma_t^0 A^{-p} T^p  \equiv C_t\cdot T^p, 
 \label{q14}
\end{equation}
where $ p=2,4,6$.
The contribution of the other channels either does not change or decreases compared to the standard situation.
The maximum contribution to the temperature dependence of the total Higgs boson creation  cross section is made by the $ M_{31}(\varepsilon)$ channel involving the $t$-quark at $p=6$ 
\begin{equation}
\sigma_t(T) = T^{6q} \sigma_t^{in}, 
 \label{q15}
\end{equation}

The results of measurements of the cross-section of the  Higgs bosons creation in four-lepton decay, obtained at the LHC for a number of years
during the collision of proton beams of different energies, are given in the next review \cite{PDG-2020},
compiled by the Particle Date Group. These data in picobarn [pb] units are reflected in the table \ref{Tab1}.
\begin{table}[H]  
\caption{Higgs boson production cross sections}
\vskip 7pt
\begin{center}
\begin{tabular}{|c|c|c|c|c|}
\hline
T, TeV & 7 & 8 & 13 & 14 \\
\hline
$\sigma_{tot}$, pb & 17 & 22 & 56 & 57 \\
\hline
\end{tabular}
\end{center}
\label{Tab1}
\end{table} 

\noindent It follows from the data in the table \ref{Tab2} that the measured  cross-sections
demonstrate a quadratic dependence on the energy of $\sigma_{tot} \sim T^2. $

\begin{table}[ht] 
\caption{Energy dependence of the Higgs boson production cross section}
\vskip 5pt
\begin{center}
\begin{tabular}{|c|c|c|c|c|}
\hline
$T_n$, TeV & $T_1$=7 & $T_2$=8 & $T_3$=13 & $T_4$=14 \\
\hline
\rowcolor[gray]{.9}
$\sigma_{tot}(T_n)/\sigma_{tot}(T_1)  $ & 1 & 1,29 & 3,29 & 3,35 \\
\hline
$ T_n/T_1 $ & 1 & 1,14 & 1,86 & 2   \\
\hline
\rowcolor[gray]{.9}
$ (T_n/T_1)^2 $  & 1 & 1,30 & 3,46 & 4 \\
\hline
$ (T_n/T_1)^3 $  & 1 & 1,69 & 6,43 & 8 \\
\hline
\end{tabular}
\end{center}
\label{Tab2}
\end{table} 

To bring into agreement
the theoretical dependence (\ref{q15}) of the cross section for the production of Higgs bosons with increasing energy (temperature) with experimental data, let us use the free parameter $q$, which defines the relationship between the contraction parameter and the temperature of the Universe, and choose $q=\frac{1}{3}$. As a result, for the $t$-quark creation  channel $ M_{31}(\varepsilon)$ (Fig.\ref{fig5}) with the largest dependence on temperature (\ref{q15}), we obtain the same quadratic dependence $\sigma_t(T) \sim T^2 $ as for the experimental cross sections $ \sigma_{tot }\sim T^2. $ Other growing cross sections are proportional to the temperature in fractional powers $ \sim T^{4/3} $ for $p=4$ and $ \sim T^{2/3} $ for $p=2$.

The cross sections for the creation  of Higgs bosons in the $ \sigma_{tot}$ four-lepton decay, measured at the LHC, are a summary result that takes into account the contribution of both $t$- and $b$-quarks, as well as all their colors (components).
Therefore, it is impossible to directly use these data to analyze the dependence of the production cross section on $T$ in different channels due to the "splitting"$\,$ of the processes of production of Higgs bosons by a loop of virtual quarks
  when taking into account the contribution of color components.
Additional assumptions are needed about the share of $t$- and $b$-loop contributions in general,
  on the contributions of each color component of quarks to the total cross section, and others.
However, a certain conclusion can be drawn that {\bf the hypothesis of contraction of the gauge group of the Standard Model does not contradict the available experimental data on the cross sections for the creation  of Higgs bosons.}


\section{Conclusion}

In gauge-type theories, a set of particles is chosen depending on the problem posed, and the gauge group determines the interaction between the particles of the model. Therefore, a change in the gauge group, in particular, its simplification with the help of contraction (transition to the limit), inevitably leads to a change in the processes occurring during the interaction of model particles.

In this paper, we analyzed the dominant mechanism for the production of Higgs bosons at the LHC.
The Feynman diagram (Fig.\ref{fig2}) of this process has been transformed to take into account the modified Lagrangian
of the  Standard model with contracted gauge group.
The right side of the diagram, which is responsible for electroweak processes, gives the amplitude $ \sim T^{-5} $ for the $t$ quark or $ \sim T^{-6} $ for the $b$ quark that decreases with increasing temperature, which is in full agreement with the results of \cite{Gr-2021}, since the interactions of particles depend on the contraction parameter and decrease along with it.
In other words, for the same number of Higgs bosons produced, the number of four-lepton events that appear decreases as the energy of the colliding particles increases.

The behavior of the left side of the diagram describing the creation   of the Higgs boson by a loop of virtual quarks does not depend on the type of quarks ($t$ or $b$) and is determined only by the strong interaction between their color components.
Thus, the total production  process is split into a number of channels, depending on the type of color components. 
The channel cross sections can either increase or decrease as $T$ increases.

The total amplitudes of the various channels for the production and registration of the Higgs boson depend on temperature according to (\ref{q13}),
and the growing sections are described by the relations (\ref{q14}), with the maximum growth given by the formula (\ref{q15}).
When choosing the exponent $q=\frac{1}{3}$, the behavior of the theoretical cross section of the process (Fig.\ref{fig5}) $\sigma_t(T) \sim T^2 $
coincides with the experimental $ \sigma_{tot} \sim T^2 $, which leads to the conclusion about the consistency
  hypotheses about the contraction of the gauge group of the Standard Model and experimental data on the cross sections for the creation  of Higgs bosons.

\vspace{5mm}

  {\it The author is grateful to V.V.~Kuratov for useful discussions and help in preparing the paper}.


\selectlanguage{english}


\begin{thebibliography}{999}
%


\bibitem{R-99}
\textit{Rubakov V.~A.} 
Classical Gauge Fields.
 Moscow: Editorial URSS, 1999 336~p. [in Russian].
%
\bibitem{Em-2007}
\textit{Emel'yanov~V.~M.}
Standard model and its extensions.
Moscow: Fizmatlit, 2007 584~p. [in Russian].


\bibitem{Gr-2016}
\textit{Gromov~N.~A.}
Elementary particles in the early Universe.
 J. Cosmol. Astropart. Phys. 2016. Vol. 03. P. 053.

\bibitem{Grom-2020}
 \textit{Gromov N.~A.}
 High-Energy Standard Model from the Gauge Group Contraction. Physics of Particles and Nuclei. 2020. Vol. 51. No. 4. P. 540--544. DOI: 10.1134/S1063779620040310. 

\bibitem{Gr-2020}
\textit{Gromov~N.~A.}
Particles in the Early Universe:
High-Energy   Limit of the Standard Model from the Contraction of Its Gauge Group.
Singapure: World Scientific, 2020 159~p.
%
\bibitem{GoR-11}
  \textit{Gorbunov D.~S.,  Rubakov V.~A.} 
 Introduction to the Theory of the Early Universe: Hot Big Bang Theory. 
 World Scientific,  2011.  
%
\bibitem{L-1990}
\textit{Linde A.~D.} 
 Particle Physics and Inflationary Cosmology.  Moscow: Nauka, 1990  280 p. [in Russian].
%
\bibitem{IW-53}
  \textit{In\"on\"u E.,  Wigner E.~P. }    
 On the contraction of groups and their representations.
 Proc. Nat. Acad. Sci. USA. 1953. Vol. 39. P. 510--524.
%
\bibitem{Gr-12}
\textit{Gromov~N.~A.}
Contractions of classical and quantum groups.
Moscow: Fizmatlit, 2012 318~p. [in Russian].
%
\bibitem{G-2020}
\textit{Gromov~N.~A.} 
    Lagrangian and Feynman diagrams of the standard model with a contracted gauge group.
Proc. of the Komi Sci. Centre,  Ural Branch, RAS. 2020.  No. 4(44). P. 16--22 [in Russian].
DOI 10.19110/1994-5655-2020-4-16-22.

\bibitem{Gr-2021}
 \textit{Gromov~N.~A.}
    Dependence of the cross sections for electroweak processes on the temperature of the Universe.
Proc. of the Komi Sci. Centre,  Ural Branch, RAS. 2021.  No. 6(52). P. 66--72 [in Russian]. 
DOI 10.19110/1994-5655-2021-6-66-72.

\bibitem{GKK-2020}
\textit{Gromov~N.~A., Kostyakov I.~V., Kuratov V.~V.} 
Diagonal contractions of small-dimension unitary algebras.
  Proc. of the Komi Sci. Centre,  Ural Branch, RAS. 2020.  No. 4(44). P. 23--29 [in Russian]. 
DOI 10.19110/1994-5655-2020-4-23-29.

\bibitem{PDG-2020}
 \textit{Zyla P.~A. et al (Particle Date Group)}
The Review of Particle Physics.
Prog. Theor. Exp. Phys. 2020. P. 083C01.

\end{thebibliography}
\end{document}